\documentclass[preprint,12pt]{elsarticle}

\usepackage{amssymb}
\usepackage{amsthm}
\usepackage{amsmath}
\usepackage{bm}
\usepackage{here}
\usepackage{color}
\usepackage{url}
\usepackage{ulem}

\journal{Mathematics and Computers in Simulation}

\begin{document}

\begin{frontmatter}



\title{Checking the Quality of Approximation of $p$-values 
  in Statistical Tests for Random Number Generators 
  by Using a Three-Level Test}


\author[haramoto]{Hiroshi Haramoto\corref{cor1}}
\address[haramoto]{Faculty of Education, Ehime University, Ehime 790-8577, Japan}
\ead{haramoto@ehime-u.ac.jp}
\cortext[cor1]{Corresponding Author}

\author[matsumoto]{Makoto Matsumoto}
\address[matsumoto]
{Graduate School of Sciences, Hiroshima University, Hiroshima 739-8526, Japan}
\ead{m-mat@math.sci.hiroshima-u.ac.jp}

\begin{abstract}
Statistical tests of pseudorandom number generators (PRNGs) are 
applicable to any type of random number generators and are
indispensable for evaluation. While several practical packages 
for statistical tests of randomness exist, they may suffer from 
a lack of reliability: for some tests, the amount of approximation error 
can be deemed significant.
Reducing this error by finding a better approximation is necessary, 
but it generally requires an enormous amount of effort. 
In this paper, we introduce an experimental method for revealing defects 
in statistical tests by using a three-level test proposed by Okutomi and Nakamura. 
In particular, we investigate the NIST test suite and the test batteries 
in TestU01, which are widely used statistical packages. 
Furthermore, we show the efficiency of several modifications for some tests.   
\end{abstract}


\end{frontmatter}


\section{Introduction}

Statistical testing of pseudorandom number generators (PRNGs) is
indispensable for their evaluation and many such test suites exist. 
Widely used examples are 
TestU01 by L'Ecuyer and Simard \cite{L'Ecuyer:2007:TCL:1268776.1268777}, 
and the test suite of the National Institute of Standards and Technology 
(NIST) \cite{Bassham:2010:SRS:2206233}.

Those suites are easy to apply to PRNGs, and further tests are still 
being designed.
However, implementers and users always face an important problem  
in determining whether each test yields correct $p$-values. 
Common problems include making tests based on incorrect mathematical 
analyses, parameter selection through experiments, 
poor implementations damaging testing credibility, etc. 
\textcolor{black}{Moreover, some statistical tests yield erroneous results
because they use approximation formulas for $p$-values with
non-negligible error. 
Therefore, checking accuracy of the approximation formula 
is important.}

The aim of this paper is to develop a method for checking the 
\textcolor{black}{quality of the approximation for the $p$-values}
of statistical tests by using a three-level test. 
This method has the merit of being easily conducted experimentally. 
Furthermore, our criterion only makes use of the uniformity of $p$-values, 
meaning that a wide range of tests can be subjected to the three-level 
method. 
Additionally, the result of this test is a $p$-value, so it is easy to 
understand as a figure of merit in statistical tests.

The rest of this paper is organized as follows. 
In Section 2, we briefly review statistical testing for PRNGs. 
In section 3, we consider a three-level test for checking 
\textcolor{black}{the quality of the approximation for the $p$-values}
of statistical tests proposed by Okutomi and Nakamura \cite{110007504717}.
In section 4, we present several results for the NIST test suite, 
SmallCrush and Crush in TestU01. 
We also present some modifications to those suites. 
These results support the usefulness of the three-level test. 

\section{Statistical testing for PRNGs \textcolor{black}{and approximation
 error}}

This section gives a brief explanation on statistical testing for PRNGs, 
especially one-level and two-level tests. 
You can find further descriptions and explanations in 
\cite{Knuth:1997:ACP:270146, rLEC92a, MR1310607, doi:10.1080/00949659708811859, 
  doi:10.1287/opre.48.2.308.12385, TestU01Manual}.
\textcolor{black}{Our aim is to use these methods to 
evaluate the precision of the approximations used
in statistical tests.}

Let $I$ denote the two element set $\{ 0, 1\}$ or 
the interval $[0, 1)$.
Let $X_1, X_2, \ldots$ be random variables distributed over $I$,
with each $X_k$ representing the $k$-th output of the tested PRNG.
A statistical test (called a one-level test) looks for empirical evidence 
against the null hypothesis
$$
\mathcal{H}_0 : X_1, X_2, \ldots, X_n \underset{{i.i.d.}}{\sim} U(I) 
$$ 
with a test statistic
$$
f :  I^n \to \mathbb{R}.
$$
 
Let $\bm{X} = (X_1, \ldots, X_n)$. 
\textcolor{black}{
In a statistical test, we assume that 
the distribution of $f(\bm{X})$ under $\mathcal{H}_0$ 
is well-approximated by a known (cumulative) distribution $F$.
} 
Thus, for our purpose to test the exactness of the 
approximation under $\mathcal{H}_0$,
we make the following hypothesis
$$
\mathcal{H}' : f(\bm{X}) {\sim} F. 
$$ 
Let $\bm{a} = (a_1,\ldots, a_n) \in I^n$ 
be an output sequence of the PRNG. If the $p$-value 
$$
F(f(\bm{a}))=\Pr \left( f(\bm{a}) \leq f(\bm{X})  \right)
$$ 
is too close to 0 or too close to 1, then 
either $\mathcal{H}_0$ or $\mathcal{H}'$ is rejected.
In usual tests for PRNG, $\mathcal{H}'$ is assumed and hence
the randomness of PRNG ($\mathcal{H}_0$) is rejected.
In this manuscript, $\mathcal{H}_0$ is assumed 
and hence the precision of the approximation ($\mathcal{H}'$)
is rejected.

If the $p$-value is very small (e.g., less than $10^{-10}$), 
then it is clear that either $\mathcal{H}_0$ or $\mathcal{H}'$ 
is rejected.
However, it is difficult to judge if the $p$-value is suspicious 
but is not very small (such as $10^{-4}$, for example). 
In order to avoid such difficulties, a two-level test is often used, 
see \cite{Knuth:1997:ACP:270146, rLEC92a}. 
A two-level test can be considered as a composite function 
$$
I^{nN} \stackrel{f^N}{\longrightarrow} 
\mathbb{R}^N \stackrel{g}{\longrightarrow} \mathbb{R}, 
$$
where $f$ is the test statistic of the one-level test 
and $f^N$ is defined by 
$$
f^N(\bm{a}_1, \ldots, \bm{a}_N) := 
\left( f(\bm{a}_1), \ldots, f(\bm{a}_N) \right) ~~~     
(\bm{a}_1, \ldots, \bm{a}_N \in I^{n}).
$$
At the second level, the function $g$ corresponds to 
a Goodness-Of-Fit (GOF) test that compares 
the empirical distribution of the $N$ $p$-values
$$
F(f(\bm{a}_1)), \ldots, F(f(\bm{a}_N))
$$
from the observations $f(\bm{a_1})$, $\ldots$, $f(\bm{a}_N)$ 
with its theoretical distribution; 
the sample size at the second level is $N$.
If the $p$-value at the second level is small, 
either $\mathcal{H}_0$ or $\mathcal{H}'$ 
is rejected.

Two-level tests permit one to apply the test with a larger total sample size 
to increase its power. 
Hence, if the generator fails the test in this particular way, 
then the $p$-value at the second level tends to become extremely small value 
as the sample size $N$ is increased. 

However, the $p$-value also tends to be very small if the approximations
of the $p$-values at the first level 
\textcolor{black}{is not good enough} (i.e.\ if ${\mathcal H}'$ fails).
In this case, computational errors accumulate at each level 
and two-level tests detect the Lack-Of-Fit of that approximation, 
leading to rejection even if the generator is good 
\cite{rLEC92a, doi:10.1287/opre.48.2.308.12385, rLEC02c, 6135498}.

\section{Checking the quality of the approximation of the $p$-value 
  by using a three-level test}

Although it is easy to extend the level of a statistical test 
from two to three (or higher) using a technique such as
$$
I^{nNN'} \,
\stackrel{f^{N \times N'}}{\longrightarrow} 
\mathbb{R}^{NN'} \stackrel{g^{N'}}{\longrightarrow}
\mathbb{R}^{N'} \stackrel{h}{\longrightarrow} \mathbb{R},  
$$
this type of test is \textcolor{black}{often useless,
because the approximation error of the second level
may destroy the result: the resulting $p$-values tend to be too close
to 0}.

By contrast, Okutomi and Nakamura proposed a three-level test that 
can be considered as reliable as a two-level one \cite{110007504717}. 
The novelty of their method is that it uses an error-free function 
at the second level. 
This allows us to increase the sample size by $N'$ times, 
and consequently to increase the power while  
avoiding an accumulation of computational errors. 
Okutomi and Nakamura originally intended to develop 
a new statistical test for PRNGs, 
but their method is useful to check the quality of the approximation of statistical tests.

Let $f$ be an $n$-variable statistic corresponding to the one-level test. 
Suppose we want to check the quality of the approximation of the
distribution of \textcolor{black}{$f(\bm{X})$}, namely $\mathcal{H}'$.
Let $(\bm{a}_1, \ldots, \bm{a}_{NN'}) \in I^{n N N'}$ 
($\bm{a}_i \in I^n$) be a sequence of $NN'$ vectors in $I^n$.
At the first level, we compute 
$f(\bm{a}_1)$, $\ldots$, $f(\bm{a}_{N N'})$.

\textcolor{black}{Here we make an assumption: the approximating distribution 
$F$ in the hypothesis ${\mathcal H}'$ is assumed 
to be continuous.}
Under ${\mathcal H}_0$ and ${\mathcal H}'$, the probability distribution of 
$F(f(\bm{X}))$ 
is uniform in $[0,1]$.
This is proved by 
$\Pr(F(f(\bm{X}))\leq p)=\Pr(f(\bm{X})\leq F^{-1}(p))
=F(F^{-1}(p))=p$, where $F^{-1}$ is the
generalized inverse distribution function
$F^{-1}(p)=\inf\{x \in \mathbb{R} \mid F(x) \geq p\}$
and the equalities follow from the continuity of $F$.
\textcolor{black}{Note that in the case of $I=\{0,1\}$, 
$f(\bm{X})$ cannot have a continuous distribution,
thus ${\mathcal H}'$ must have some error.
Therefore, we should distinguish the right and left $p$-values 
\cite{L'Ecuyer:2007:TCL:1268776.1268777, TestU01Manual}. 
In this paper, the assumption ${\mathcal H}'$
means that $F$ is an approximation good enough
so that the statistical tests behave well.
Thus, ${\mathcal H}'$ includes 
the assumption that each probability mass is small 
enough to be negligible by itself. 
}

We fix an arbitrary significance level $\alpha \in (0, 1)$.
The function $g$, which corresponds to the second level,
is the function that counts the number $T_i$ of $p$-values 
greater than or equal to $\alpha$ in 
$$
F(f(\bm{a}_{1+(i-1)N})), \ldots,
F(f(\bm{a}_{iN}))
$$
for $i=1, \ldots, N'$.
Under the hypotheses
$\mathcal{H}_0$, $\mathcal{H}'$ and the continuity
of $F$, 
the distribution of the above $p$-values should be 
independently uniformly distributed over the interval 
$[0,1]$, as shown above.
Therefore, $T_i$ should have the binomial distribution 
$B(N, 1-\alpha)$.

Finally, at the third level, we compare the empirical distributions of 
$T_1$, $\ldots$, $T_{N'}$ and $B(N, 1-\alpha)$ via a GOF test. 
If the resulting $p$-value at the third level is extremely small, 
\textcolor{black}{it strongly suggests that either ${\mathcal H}_0$ or
${\mathcal H}'$ fails. In our purpose, we use good
PRNGs so that ${\mathcal H}_0$ is assumed,
and consider that ${\mathcal H}'$ is rejected, or equivalently
the approximation of $f(\bm{X})$ by $F$ is not good enough.}

In this paper, following \cite{110007504717}, we use the parameters 
$\alpha = 0.01$, $N=10^3$, and $N'=10^3$, 
as well as the following categorization:
\begin{align*}
C_0 &= \{0, 1, \ldots, 981\}, \\
C_i &= \{981+i\} ~~~ (i=1,2,\ldots, 15), \\
C_{16} &= \{997, 998, 999, 1000\}.
\end{align*}

Let $Y_i := \# \{ T_j \mid j=1, \ldots, N', T_j \in C_i\}$ 
for $i=0, \ldots, 16$. 
We compute the $\chi^2$-value
$$
h(T_1, \ldots, T_{N'}) := \sum_{i=0}^{16} \frac{(Y_i-N'p_i)^2}{N'p_i},
$$
where $p_i = \sum_{j \in C_i} \binom{N}{j}/2^N$.
The distribution of this statistic under $\mathcal{H}_0$ and
$\mathcal{H}'$
is approximated by the $\chi^2$-distribution with 
$16$ degrees of freedom. 

\textcolor{black}{
It might seem to be more natural to 
use a GOF test on the entire distribution 
of the $N'$ $p$-values under the uniform distribution hypothesis,
by using a test such as a Kolmogorov-Smirnov test.
However, if the distribution of the test statistic at the second level
is only approximated and the approximation error is significant, 
a three-level test will detect this approximation
error, and tends to give $p$-values nearly zero.} 

On the other hand, the presented method counts only the number of 
$p$-values at the second level, which has \textcolor{black}{no} 
approximation error introduced at this level
(under the hypotheses ${\mathcal H}_0$ and
${\mathcal H}'$).
This is a reason why the proposed test is better. 

\section{Experimental results}

This section shows the experimental results of 
\textcolor{black}{the three-level test} for 
the NIST test suite, SmallCrush, and Crush in TestU01. 
In order to mimic truly random number sequences at the first level, 
we adopt Mersenne Twister (MT) \cite{DBLP:journals/tomacs/MatsumotoN98} 
and a PRNG from the SHA1 algorithm.
Note that MT fails certain tests (e.g. the linear complexity test, 
the binary matrix rank test with large matrix size) even when 
the $p$-value is computed correctly with no significant error.
Thus we need both generators in the following experiments.

\subsection{Results of the NIST test suite}

The NIST test suite 
consists of 15 statistical tests for randomness of bit sequences, 
and its result is a list of $188$ $p$-values.
Since it was published in 2001, many modifications and corrections have been studied.
However, the latest version 2.1.2 \cite{Bassham:2010:SRS:2206233}, 
released in 2014, does not incorporate several modifications.

We will show that the three-level method can reveal known defects of some statistical tests
and show the effectiveness of several proposals to increase the reliability of those tests.
In addition, 
\textcolor{black}{through the three-level methods, we deduce new
constraints}
for the Random Excursions test and the Random Excursions Variant test. 

First, we consider all tests other than the Random Excursions test and 
the Random Excursions Variant test. 
In this experiment, the sample size at the first level $n$ 
is fixed to $10^6$, as recommended by NIST.
\textcolor{black}{Recall that throughout the experiments, 
the number of iterations $N$ in the second level 
and $N'$ in the third level are both $1000$
with categorizations described in the previous section.
}

Table 1 shows the 
results 
\textcolor{black}{of the three-level test} 
for the original NIST test suite and the modified tests explained later.
The Cumulative Sums test and the Serial test have 
two statistics respectively, with two $p$-values written in Table 1. 
The Non-Overlapping Template Matching test reports $148$ $p$-values, 
thus the passing rates are filled in the table.

From our experiments, the $p$-values from the 
Discrete Fourier Transform (DFT) test, 
the Overlapping Template Matching test, 
and the Maurer's Universal Statistical test in the original test suite are 
much too small. 
Additionally, the $p$-values of the Longest Runs of Ones in a Block test 
are relatively small. 
This result indicates that those tests have some flaws.
After we applied appropriate modifications, 
the three-level test reported reasonable $p$-values for the four tests
\textcolor{black}{as described in the two columns at the right
in Table~1}.


\begin{table}[htpb]
  \begin{center}
    \caption{
      Results 
      \textcolor{black}{of the three-level test} 
      for the NIST test suite with $n = 10^6$}
    \begin{tabular}{c|c|c|c|c}
      \multicolumn{1}{l|}{} & \multicolumn{2}{|c|}{$p$-value(Original)}
      & \multicolumn{2}{|l}{$p$-value(Modified)} \\ \cline{2-5}
      Test Name & MT & SHA1 & MT & SHA1 \\ \hline 
      Frequency & $0.85$ & $0.59$ & - & - \\
      Frequency test within a Block & $0.017$ & $0.68$ & - & - \\
      Cumulative Sums Test & $0.13$, $0.64$ & $0.37$, $0.43$ & - & - \\
      Runs & $0.56$ & $0.47$ & - & - \\
      Longest Run of Ones in a Block & 3.9E$-$5 & 1.3E$-$8 & $0.44$ & $0.0011$ \\
      Binary Matrix Rank &  $0.30$ & $0.13$ & - & - \\
      Discrete Fourier Transform & 4.1E$-119$ & 7.2E$-$116 & $0.19$ & $0.026$ \\
      Non-Overlapping Template Matching & 148/148 & 148/148 & - & - \\
      Overlapping Template Matching & 7.5E$-$80 & 5.6E$-$73 & $0.70$ & $0.88$ \\
      Maurer's Universal Statistical & 8.7E$-$76 & 4.1E$-$66 & $0.99$ & $0.77$ \\
      Approximate Entropy  & $0.40$ & $0.036$ & - & - \\
      Serial & $0.67$, $0.70$ & $0.28$, $0.39$ & - & - \\
      Linear Complexity & $0.023$ & $0.0030$ & - & - 
    \end{tabular}
  \end{center}
\end{table}


Note that the current implementation of the NIST test suite 
uses one-level or two-level tests, differently from the above 
experiments, and the approximation error in $p$-values 
provided by those tests are not large.
For example, Table 2 and Table 3 show $p$-values provided by 
the NIST test suite.

\begin{table}[H]
  \begin{center}
    \caption{$p$-values of one-level tests and a two-level test for MT}
    \begin{tabular}{c|c|c|c|c|c||c}
      \multicolumn{1}{c|}{} & \multicolumn{5}{|c||}{first level ($n=10^6$)}
      & second level \\ \cline{2-6}
      Test Name & 1st & 2nd & 3rd & 4th & 5th & ($N=10^3$) \\ \hline
      Longest Run of Ones in a Block & 
      0.15 & 0.39 & 0.64 & 0.029 & 0.47 & 0.88 \\ \hline
      Discrete Fourier Transform 
      & 0.48 & 0.44 & 0.31 & 0.89 & 0.66 & 0.41 \\ \hline
      Overlapping Template Matching 
      & 0.58 & 0.69 & 0.18 & 0.47 & 0.99 & 0.15 \\ \hline
      Maurer's Universal Statistical 
      & 0.78 & 0.96 & 0.083 & 0.40 & 0.38 & 0.99
    \end{tabular}


    \caption{$p$-values of one-level tests and a two-level test for SHA1}
    \begin{tabular}{c|c|c|c|c|c||c}
      \multicolumn{1}{c|}{} & \multicolumn{5}{|c||}{first level ($n=10^6$)}
      & second level \\ \cline{2-6}
      Test Name & 1st & 2nd & 3rd & 4th & 5th & ($N=10^3$) \\ \hline
      Longest Run of Ones in a Block 
      & 0.65 & 0.50 & 0.69 & 0.44 & 0.052 & 0.64 \\ \hline
      Discrete Fourier Transform 
      & 0.73 & 0.038 & 0.13 & 0.77 & 0.34 & 0.034 \\ \hline
      Overlapping Template Matching 
      & 0.21 & 0.75 & 0.91 & 0.087 & 0.76 & 0.14 \\ \hline
      Maurer's Universal Statistical 
      & 0.32 & 0.33 & 0.63 & 0.89 & 0.090 & 0.083
    \end{tabular}
  \end{center}
\end{table}

Let us explain the modifications.
We begin by considering the DFT test. 
Let $X_k$ be the $k$-th bit of the tested sequence. 
The DFT test computes the discrete Fourier coefficients 
$$
F_i = \sum_{k=0}^{n-1} (2X_k-1) \exp(-2\pi \sqrt{-1} k i/ n), ~~~ 
i=0,1, \ldots, n/2-1. 
$$
The $p$-value of the DFT test is \textcolor{black}{approximated by}
$$
\Pr ((o_h-0.95 n/2)/\sqrt{0.05 \cdot 0.95 n / d} < Z), 
~~~ Z \sim N(0, 1)
$$
for a realization $o_h$ of the number $O_h$ of $|F_j|$'s that 
are smaller than some constant $h$. 
The latest version of the NIST test suite uses the parameter $d=4$ 
proposed by Kim et al. \cite{journals/iacr/KimUH04}. 
Subsequently, Pareschi et al. \cite{6135498} proposed $d=3.8$ 
for $n \approx 10^6$, which we use here as modification. 

The Overlapping Template Matching test uses a $\chi^2$ GOF test that
compares the empirical distribution of occurrences of a certain 
bit template with the theoretical one. NIST once used the probabilities 
derived by an approximation formula, and now it adopts more accurate values 
derived by \cite{Hamano:2007:COT:1521215.1521232}. However,  
the C-code \texttt{overlappingTemplateMatching.c} changes the new values 
to the former wrong ones. We thus remove this instruction
from the original code (lines 40--44),
\textcolor{black}{which is the modification}.  

The Maurer's Universal Statistical test detects whether the sequence can be 
significantly compressed without loss of information. 
The original test adopts an asymptotic measure. We use the 
modification by Coron \cite{Coron1999}, a variant test statistic
which enables better detection of defects in the tested sequence. 

The Longest Runs of Ones in a Block test also uses a $\chi^2$ GOF test. 
The NIST test suite uses approximation values to four decimal places
instead of the theoretical probabilities. 
We \textcolor{black}{modify these values by more accurate} ones to fifteen decimal places.

Unlike the other tests, the Random Excursions test and 
the Random Excursions Variant test do not always yield $p$-values. 
We review the algorithms of those tests and explain why this happens.

Both tests are based on considering successive sums of the bits 
as a one-dimensional random walk. 
Let $X_1, \ldots, X_n$ be random variables distributed over $\{0,1\}$. 
The Random Excursions and Random Excursions Variant tests 
compute the partial sums
$$
S_i := \sum_{k=1}^{i} (2X_k-1), ~~~ i=1, \ldots, n,
$$
\textcolor{black}{called the $i$-th \textit{state} of the random walk.
For an integer $x$, we say that $S_i$ takes the value $x$
if $S_i=x$.
Consider the sequence $(0, S_1, \ldots, S_n, 0)$,  
and let $J$ be the number of $0$'s minus one in this sequence.}
\textcolor{black}{We call a subsequence of $(0, S_1, \ldots, S_n, 0)$
a \textit{cycle} if it has length no less than two,
it starts with $0$, ends with $0$, and contains
no $0$ between the first $0$ and the last $0$.
Hence $J$ is the total number of cycles in $(0, S_1, \ldots, S_n, 0)$.}
\textcolor{black}{
Let $x$ be an integer among 
$x=\pm 1$, $\pm 2$, $\pm 3$, $\pm 4$.
For each $x$ among the eight, the Random Excursions test uses
the test statistic consisting
of six integers $\nu_k(x)$ ($k=0,1,2,3,4,5$).
For $k<5$, $\nu_k(x)$ is the number of cycles in which
the frequency of the value $x$ in the states is exactly $k$.
For $k=5$, $\nu_5(x)$ is the number of cycles in which
the frequency of the value $x$ is $5$ or more. 
Thus, $\sum_{k=0}^5\nu_k(x)=J$ holds.
}
\textcolor{black}{The corresponding $\chi^2$ statistic is} 
$$
\chi^2 := \sum_{k=0}^5 \frac{(\nu_k(x)-J \pi_k(x))^2}{J\pi_k(x)}, 
$$
where $\pi_k(x)$ is the probability 
\textcolor{black}{that the state $S_i$ visits the value $x$ exactly $k$ times in a cycle, 
  under $\mathcal{H}_0$.}
\textcolor{black}{For the test statistic to have approximately a chi-square distribution,} 
the expectation $J\pi_k(x)$ for each $k$ should not be too small, say $J\pi_k(x) \geq 5$. 
The NIST test suite \textcolor{black}{discards the sample} if $J < 500$ 
because the minimum value of $\pi_k(x)$'s is $\pi_4(4) \approx 0.0105$.
\textcolor{black}{Thus, each test yields eight $p$-values (one for each $x$)
when $J\geq 500$, 
and yields no result when $J<500$}.

The Random Excursions Variant test computes the number $\xi(x)$ of times 
that $x$ occurs across all $J$ cycles for $x=\pm 1, \pm2, \ldots, \pm 9$.
The limiting distribution of $\xi(x)$ is known to be normal with 
mean $J$ and variance $J(4|x|-2)$ for each $x$: thus, the test suite uses 
the statistic 
$$
Z:=(\xi(x)-J)/(\sqrt{J(4|x|-2)}).
$$
The constraint is also $J \geq 500$.



\textcolor{black}{In the Random Excursions test and 
the Random Excursions Variant test, $J$ is the sample size 
in computing $p$-values.
Hence, the approximations of the statistics of the tests
by a chi-square distribution and a normal distribution 
are getting
better when the number $J$ is increased.}

\textcolor{black}
{
Since these tests discard some parts of the output 
of PRNG, the formalism of the three-level test
does not apply as it is. However, for the both tests,
the first level procedure yields a sequence
of $p$-values which are uniform i.i.d in $[0,1]$
under the hypotheses ${\mathcal H}_0$ and ${\mathcal H}'$.
We iterate the first-level tests until we obtain 
$N(=1000)$ sample $p$-values.
Then, the rest of the three-level test works in the
same manner.
}

We show the results of the three-level test for the Random Excursions test 
in Table 4, and for the Random Excursions Variant test in Table 5.
Note that we use the sample size at the first level 
$n=10^7$ to decrease the number of tests in which the test procedure
is discontinued. 

\begin{table}[h]
  \begin{center}
    \caption{$p$-values of 
      \textcolor{black}{the three-level test} 
      of the Random Excursions test}
    \begin{tabular}{c|c|c|c|c|c|c|c|c}
      \multicolumn{1}{c|}{} & \multicolumn{2}{|c|}{$J\geq 500$}
      & \multicolumn{2}{|c}{$J\geq 1000$} & \multicolumn{2}{|c}{$J\geq 1500$}
      & \multicolumn{2}{|c}{$J\geq 2000$}\\ \cline{2-9}
      $x$ & MT & SHA1 & MT & SHA1 & MT & SHA1 & MT & SHA1 \\ \hline 
      $-4$ & 1.0E$-$10 & 9.1E$-$20 & 2.7E$-$03 & 4.9E$-$14 & 2.6E$-$02 & 6.5E$-$11 &
      1.1E$-$03 & 1.2E$-$04 \\
      $-3$ & 3.2E$-$06 & 8.1E$-$07 & 3.4E$-$04 & 2.3E$-$02 & 1.2E$-$01 & 2.4E$-$01 &
      8.4E$-$02 & 4.1E$-$01 \\
      $-2$ & 4.5E$-$01 & 3.1E$-$01 & 3.8E$-$01 & 1.9E$-$01 & 3.1E$-$01 & 2.1E$-$04 &
      3.2E$-$01 & 1.1E$-$01 \\
      $-1$ & 6.4E$-$02 & 8.9E$-$01 & 8.6E$-$01 & 2.9E$-$01 & 1.6E$-$01 & 4.1E$-$01 &
      2.8E$-$02 & 2.8E$-$01 \\
      $1$  & 9.0E$-$02 & 6.3E$-$01 & 5.7E$-$01 & 3.5E$-$01 & 2.9E$-$01 & 5.8E$-$01 &
      2.5E$-$01 & 4.7E$-$01 \\
      $2$  & 3.8E$-$02 & 5.8E$-$02 & 4.0E$-$02 & 1.3E$-$01 & 9.2E$-$03 & 4.7E$-$01 &
      8.2E$-$02 & 1.7E$-$01 \\
      $3$  & 3.5E$-$06 & 2.5E$-$08 & 8.5E$-$04 & 6.8E$-$05 & 1.5E$-$03 & 1.6E$-$04 &
      5.5E$-$03 & 9.1E$-$05 \\
      $4$  & 6.7E$-$16 & 4.7E$-$18 & 5.7E$-$03 & 6.0E$-$10 & 9.7E$-$02 & 4.8E$-$05 &
      5.3E$-$02 & 4.0E$-$06
    \end{tabular}
  \end{center}
\end{table}

\begin{table}[h]
  \begin{center}
    \caption{$p$-values of the Random Excursions Variant test}
    \begin{tabular}{c|c|c|c|c}
      \multicolumn{1}{c|}{} & \multicolumn{2}{|c|}{$J\geq 500$}
      & \multicolumn{2}{|c}{$J\geq 1000$} \\ \cline{2-5}
      $x$ & MT & SHA1 & MT & SHA1 \\ \hline
      $-9$ & 1.5E$-$07 & 9.3E$-$10 & 2.5E$-$01 & 5.3E$-$01 \\
      $-8$ & 2.8E$-$07 & 3.6E$-$05 & 3.3E$-$01 & 8.1E$-$01 \\
      $-7$ & 1.3E$-$08 & 2.2E$-$05 & 3.9E$-$01 & 2.4E$-$01 \\
      $-6$ & 8.9E$-$03 & 5.9E$-$03 & 6.0E$-$01 & 6.9E$-$01 \\
      $-5$ & 5.1E$-$02 & 1.6E$-$02 & 9.2E$-$01 & 7.9E$-$01 \\
      $-4$ & 7.4E$-$04 & 1.7E$-$01 & 4.5E$-$01 & 6.6E$-$01 \\
      $-3$ & 8.6E$-$03 & 4.7E$-$03 & 5.2E$-$01 & 8.7E$-$01 \\
      $-2$ & 2.5E$-$02 & 8.3E$-$01 & 3.1E$-$01 & 1.2E$-$02 \\
      $-1$ & 4.9E$-$01 & 7.1E$-$04 & 7.2E$-$01 & 9.3E$-$01 \\
      $1$ & 3.4E$-$01 & 8.8E$-$01 & 2.0E$-$01 & 7.0E$-$01 \\
      $2$ & 3.4E$-$02 & 1.6E$-$02 & 6.1E$-$02 & 1.3E$-$01 \\
      $3$ & 6.0E$-$01 & 1.6E$-$03 & 7.0E$-$01 & 8.5E$-$01 \\
      $4$ & 8.5E$-$02 & 1.4E$-$02 & 2.1E$-$01 & 2.5E$-$01 \\
      $5$ & 1.5E$-$01 & 1.5E$-$03 & 5.5E$-$01 & 5.6E$-$01 \\
      $6$ & 1.8E$-$03 & 5.5E$-$05 & 1.5E$-$01 & 1.5E$-$01 \\
      $7$ & 2.8E$-$03 & 1.0E$-$06 & 6.7E$-$01 & 5.3E$-$01 \\
      $8$ & 2.7E$-$04 & 1.0E$-$05 & 8.7E$-$02 & 9.0E$-$01 \\
      $9$ & 1.2E$-$07 & 2.6E$-$09 & 5.5E$-$01 & 4.7E$-$01
    \end{tabular}
  \end{center}
\end{table}

From our experiments, the Random Excursions test 
\textcolor{black}{for $x=4$ shows some flaw up to 
$J=1500$. For the safety, we recommend a stronger constraint
$J\geq 2000$ than $J\geq 500$ which NIST specified, with
a larger sample size $n=10^7$. 
For the Random Excursions Variant test, from the too small 
$p$-values for $x=\pm 9$, we recommend a constraint
$J \geq 1000$.}
\subsection{Results for SmallCrush and Crush in TestU01}

We examine the 
\textcolor{black}{quality of the approximation of the $p$-values} 
of SmallCrush and Crush batteries in TestU01. 
SmallCrush battery consists of $10$ statistical tests ($16$ statistics).
Of those tests, the \texttt{smarsa\char`_BirthdaySpacings} test and 
one of the \texttt{sknuth\char`_Collision} tests are based on a Poisson distribution, 
meaning that their distributions of $p$-values are not uniform. 
We thus assess the 
\textcolor{black}{quality of the approximation of the $p$-values} 
of the remaining $14$ test statistics.
Table 6 indicates that all 14 tests 
\textcolor{black}{have the approximations of $p$-values 
  which are sufficiently accurate.}

\begin{table}[h]
  \begin{center}
    \caption{$p$-values of the 
      \textcolor{black}{three-level} test of the SmallCrush}
    \begin{tabular}{c|c|l|l}
      test name & distribution & MT &  SHA1 \\ \hline 
      \texttt{sknuth\char`_Collision} & normal &$0.057$ & $0.76$  \\
      \texttt{sknuth\char`_Gap} & $\chi^2$ & $0.059$ & $0.37$  \\
      \texttt{sknuth\char`_SimplePoker} & $\chi^2$ & $0.47$ & $0.75$  \\
      \texttt{sknuth\char`_CouponCollector} & $\chi^2$ & $0.62$ & $0.94$  \\
      \texttt{sknuth\char`_MaxOft} & normal & $0.0047$ & $0.47$  \\
      \texttt{sknuth\char`_MaxOft} & $\chi^2$ & $0.017$ & $0.62$  \\
      \texttt{svaria\char`_WeightDistrib} & $\chi^2$ & $0.50$ & $0.049$  \\
      \texttt{smarsa\char`_MatrixRank} & $\chi^2$ & $0.29$ & $0.90$  \\
      \texttt{sstring\char`_HammingIndep} & normal & $0.019$ & $0.095$  \\
      \texttt{swalk\char`_RandomWalk1} (\texttt{H}) & $\chi^2$ & $0.0020$ & $0.043$  \\
      \texttt{swalk\char`_RandomWalk1} (\texttt{M}) & $\chi^2$ & $0.011$ & $0.48$  \\
      \texttt{swalk\char`_RandomWalk1} (\texttt{J}) & $\chi^2$ & $0.26$ & $0.90$  \\
      \texttt{swalk\char`_RandomWalk1} (\texttt{R}) & $\chi^2$ & $0.83$ & $0.12$  \\
      \texttt{swalk\char`_RandomWalk1} (\texttt{C}) & $\chi^2$ & $0.23$ & $0.40$  
    \end{tabular}
  \end{center}
\end{table}

Crush battery consists of 96 tests and reports 144 $p$-values. 
We check the quality of the approximation of the 76 tests (90 statistics), 
whose statistics have continuous distributions, 
ignoring those whose statistics are discrete, namely
the \texttt{smarsa\char`_CollisionOver} test (No.3--10), 
the \texttt{smarsa\char`_BirthdaySpacings} test (No.11--17), 
the \texttt{snpair\char`_ClosePairs} test (No.18--20), 
the \texttt{snpair\char`_ClosePairsBitMatch} test (No.21--22), 
and one of the test statistics of the 
\texttt{sknuth\char`_CollisionPermut} test (No.39--40), 
where the numbers correspond to the enumeration of the tests 
in the user's guidebook \cite{TestU01Manual}.

To reduce the computation time, we check 
Crush using the following procedure.
We 
\textcolor{black}{apply the three-level test with Mersenne Twister to}
each test 
If the $p$-value is smaller than $10^{-10}$, 
we check the test with a PRNG from SHA1. 
The test (i.e. hypothesis ${\mathcal H}'$)
will be rejected if both $p$-values are smaller 
than $10^{-10}$. 

Table 7 shows the tests rejected by both Mersenne Twister and SHA1. 
Because the \texttt{sspectral\char`_Fourier3} test has three statistics,  
three corresponding $p$-values are listed in the right-most column
in Table~7,
each of which is smaller than $10^{-300}$.
The \texttt{sstring\char`_Run} test has two statistics, 
thus we listed two $p$-values in the table.
Similarly to the case of the NIST test suite,
the approximation error in the $p$-value by TestU01
is not that large even if we find $\varepsilon$ values
in these three-level tests.

\begin{table}[ht]
  \begin{center}
    \caption{Rejected tests in Crush and their $p$-values 
    ($\varepsilon$ : the $p$-value $<10^{-300}$)}
    \begin{tabular}{c|c|c|c}
      test name & parameters & MT & SHA1 \\ \hline 
      \texttt{svaria\char`_SampleCorr} & $n=5 \times 10^8$, $k=1$ & 1.8E$-$222 & 5.5E$-$237 \\
      \texttt{smarsa\char`_Savir2} & $n=2 \times 10^7$, $m=2^{20}$, $t=30$ 
      & 2.7E$-$49 & 9.9E$-$32 \\
      \texttt{scomp\char`_LempelZiv} & $n=2^{25}$ & 
      $\varepsilon$ & $\varepsilon$ \\
      \texttt{sspectral\char`_Fourier3} & $n=2^{14} \times 50000$ & 
      $\varepsilon$, $\varepsilon$, $\varepsilon$ & 
      $\varepsilon$, $\varepsilon$, $\varepsilon$ \\
      \texttt{sstring\char`_Run} & $n = 10^9$ & 
      $\varepsilon$, $\varepsilon$ & $\varepsilon$, $\varepsilon$ \\
    \end{tabular}
  \end{center}
\end{table}

\textcolor{black}{
Among these rejected tests, we find that two of them can be modified
to pass the three-level test. 
These are the \texttt{svaria\char`_SampleCorr} test
and the \texttt{sstring\char`_Run} test.
The improvements are shown in Table~8.
}
The \texttt{svaria\char`_SampleCorr} test computes a correlation between 
$X_1, \ldots, X_n$ which are random variables distributed over $[0, 1)$.
TestU01 assumes that the statistic
$$
\frac1{n-k} \sum_{j=1}^{n-k} (X_jX_{j+k}-1/4), ~~~ 
$$
has the normal distribution with mean $0$ and variance $1/{12(n-k)}$. 
Fishman \cite{Fishman:1978:PDE:539984} shows that the statistic
$$
\frac1{n-k} \sum_{j=1}^{n-k} (X_j-1/2) (X_{j+k}-1/2)
$$
converges to normal with mean $0$ and variance $1/{144(n-k)}$. 
\textcolor{black}{We modified the original statistic
$\frac1{n-k} \sum_{j=1}^{n-k} (X_jX_{j+k}-1/4)$ to 
$\frac1{n-k} \sum_{j=1}^{n-k} (X_j-1/2) (X_{j+k}-1/2)$.
}

The \texttt{sstring\char`_Run} test is a variant of the run test applicable
to a bit sequence, which yields two $p$-values: 
the test statistics are 
based on a normal distribution and a $\chi^2$ distribution.

Let $Y$ be the total number of bits needed to obtain $2n$ runs.
Under $\mathcal{H}_0$, we have $Y = \sum_{i=1}^{2n} X_i +2n$
where $X_i$ are independent geometric random variables with parameter $1/2$.
TestU01 adopts the statistic $(Y-4n)/\sqrt{8n}$ and assumes 
that it can be approximated by the standard normal distribution.
However, the expectation of $X_i$ is $1$ and the variance 
of $X_i$ is $2$, so
$$
E[Y] = \sum_{i=1}^{2n} E[X_i] + 2n = 4n,  
~~~ V[Y] = \sum_{i=1}^{2n} V[X_i] = 4n.
$$
Thus the appropriate statistic is $(Y-4n)/\sqrt{4n}$, this is the 
modification.

The other test statistic is 
$$
\sum_{i=1}^{k}\dfrac{(X_{0,i}-np_i)^2}{np_i(1-p_i)} + 
\sum_{i=1}^{k}\dfrac{(X_{1,i}-np_i)^2}{np_i(1-p_i)},
$$
where $X_{0, i}$ and $X_{1, i}$ are the number of runs of $0$'s and $1$'s 
of length $i$ for $i=1, \ldots, k$, where $k$ is some positive integer, and
$p_i=2^{-i}$. TestU01 assumes that the statistic has approximately 
the $\chi^2$ distribution with $2(n-1)$ degrees of freedom for 
a $\chi^2$ GOF test.
However, the factor $1-p_i$ in the denominator seems to be unnecessary,
so our modification removes them.  

Table 8 shows the $p$-values for those test statistics. 
The results indicate that the above modifications are satisfactory
in improving the reliability of the tests.

\begin{table}[ht]
  \begin{center}
    \caption{$p$-values of the original tests and their modifications 
      ($\varepsilon$ : the $p$-value $<10^{-300}$)}
    \begin{tabular}{c|c|c|c|c}
    \multicolumn{1}{l|}{} & \multicolumn{2}{|c|}{$p$-value (Original)}
    & \multicolumn{2}{|l}{$p$-value(Improved)} \\ \cline{2-5}
    test name & MT & SHA1 & MT & SHA1 \\ \hline 
    \texttt{svaria\char`_SampleCorr} & 1.8E$-$222 & 5.5E$-$237
    & $0.498$ & $0.825$ \\ 
    \texttt{sstring\char`_Run} (normal) & $\varepsilon$ & $\varepsilon$
    & $0.657$ & $0.302$ \\ 
    \texttt{sstring\char`_Run} (chi-squared) & $\varepsilon$ & $\varepsilon$
    & $0.715$ & $0.0479$
    \end{tabular}
  \end{center}
\end{table}

\textcolor{black}{We discuss on the rest three tests, for which we are not 
able to give satisfactory modifications.}
The \texttt{smarsa\char`_Savir2} test is a modified version of 
the Savir test proposed by Marsaglia \cite{1676623w}. 
Let $U_1$, $U_2$, $\ldots$, $U_t$ be independent uniform random variables 
over $(0, 1)$. For a given $m$, the random integers 
$I_1$, $I_2$, $\ldots$, $I_t$ are defined by 
$I_1 = \lceil m U_1 \rceil$, $I_2 = \lceil I_1 U_2 \rceil$, $\ldots$, 
$I_t = \lceil I_{t-1} U_t \rceil$. It thus generates $n$ values of $I_t$ 
and compares their empirical distribution with the theoretical one  
via a $\chi^2$-test. 

TestU01 recommends the values of $m$ and $t$ that satisfy 
$m \approx 2^t$ and Crush adopts $m=2^{20}$ and $t=30$.
Table 9 shows the $p$-values obtained with $n=2 \times 10^7$, $m=2^{20}$
and various values of $t$.

The $p$-values are slightly suspicious but not too small. 
Therefore, it is necessary to investigate the tests mathematically,
but we are not able to manage this at present.
Tentatively, we propose to take $t=9$ for a compromise between 
the reliability of the test and choosing a larger value of $t$.


\begin{table}[H]
  \begin{center}
    \caption{$p$-values of the \texttt{smarsa\char`_Savir2} test 
      for various $t$'s}
    \begin{tabular}{c}
      \begin{minipage}{0.5\hsize}
        \begin{center}
          \begin{tabular}{c|c|c}
            $t$ & MT & SHA1 \\ \hline
            $5$ & 2.7E$-$07 & 6.9E$-$06 \\
            $6$ & 4.2E$-$12 & 3.4E$-$05 \\
            $7$ & 2.3E$-$10 & 3.5E$-$08 \\
            $8$ & 5.2E$-$06 & 1.5E$-$09 \\
            $9$ & 1.1E$-$06 & 1.6E$-$05 \\
            $10$ & 3.2E$-$04 & 2.1E$-$13 \\
            $11$ & 5.6E$-$14 & 1.6E$-$15 
          \end{tabular}
        \end{center}
      \end{minipage}
      \begin{minipage}{0.5\hsize}
        \begin{center}
          \begin{tabular}{c|c|c}
            $t$ & MT & SHA1 \\ \hline
            $12$ & 1.9E$-$06 & 1.6E$-$12 \\
            $13$ & 1.6E$-$07 & 2.3E$-$08 \\
            $14$ & 3.8E$-$06 & 4.9E$-$16 \\
            $15$ & 4.7E$-$16 & 1.7E$-$17 \\
            $20$ & 2.2E$-$13 & 2.0E$-$18 \\
            $25$ & 1.3E$-$28 & 1.2E$-$17 \\
            $30$ & 2.7E$-$49 & 9.9E$-$32 
          \end{tabular}
        \end{center}
      \end{minipage}
    \end{tabular}
  \end{center}
\end{table}

The \texttt{scomp\char`_LempelZiv} test measures the compressibility of 
the bit sequence using the Lempel-Ziv compression algorithm. TestU01 uses 
approximations of the mean and variance obtained by simulation.  
The \texttt{sspectral\char`_Fourier3} test is a kind of DFT tests 
proposed by Erdmann. 
However, the authors of TestU01 claim that those tests tend not 
to be very sensitive. 
Indeed, the resulting $p$-values of those tests 
are smaller than $10^{-300}$, so more mathematical justifications 
for those tests are needed.

\section{Concluding remarks}
We introduced a three-level test to check the 
\textcolor{black}{the quality of the approximation for the $p$-values} 
in statistical tests 
for PRNGs. 
\textcolor{black}{We find that some statistical tests use approximation
with some flaw. We list some of such tests from NIST and TestU01.
This does not mean that these tests are erroneous,
but the reliability of the tests is increased if the approximation
is improved. We give three satisfactory modifications to three tests 
in Crush, and propose new parameters for several tests from this viewpoint.}
\textcolor{black}{In this study, 
we need to assume that the approximated statistics
are continuous, because our three-level test is based on
the uniformity of $p$-values in $[0,1]$ at the first level.}
This condition is 
not essential: if the distribution of $p$-values can be computed exactly, 
we can conduct the three-level test with an appropriate GOF test 
at the third level. For example, 
the exact probability formula of the \texttt{smarsa\char`_BirthdaySpacings} 
test is presented in \cite{Knuth:1997:ACP:270146}. 
It indicates the possibility 
of calculating the exact distribution of its $p$-values. 
In future work, we hope to assess the reliability of all of 
the remaining tests in Crush battery.

According to the original proposal presented in \cite{110007504717}, 
we employ a $\chi^2$ GOF test at the third level. 
However, a Kolmogorov-Smirnov (KS) test seems to be more appropriate 
and more powerful. 
An accurate approximation of the KS distribution is now available 
\cite{JSSv039i11}, so we should experiment with this method 
to obtain more decisive conclusions.

\section*{Acknowledgement}
The computation in this work has been done using
the facilities of the Institute of Statistical Mathematics.
The authors would like to express their sincere gratitude 
to anonymous referees for their invaluable comments and 
Professor Pierre L'Ecuyer for his encouragements. 
This work was supported by JSPS KAKENHI Grant 
Numbers 15K13460, 26310211, 16K13750, 17K14234, and 18K03213. 
This work was also supported by JST CREST.



\section*{References}
\bibliographystyle{model1b-num-names.bst} 
\bibliography{haramoto01}

\begin{thebibliography}{18}
\expandafter\ifx\csname natexlab\endcsname\relax\def\natexlab#1{#1}\fi
\providecommand{\bibinfo}[2]{#2}
\ifx\xfnm\relax \def\xfnm[#1]{\unskip,\space#1}\fi
\bibitem[{Bassham et~al.(2010)Bassham, Rukhin, Soto, Nechvatal, Smid, Barker,
  Leigh, Levenson, Vangel, Banks, Heckert, Dray and
  Vo}]{Bassham:2010:SRS:2206233}
\bibinfo{author}{L.E. Bassham, III}, \bibinfo{author}{A.L. Rukhin},
  \bibinfo{author}{J.~Soto}, \bibinfo{author}{J.R. Nechvatal},
  \bibinfo{author}{M.E. Smid}, \bibinfo{author}{E.B. Barker},
  \bibinfo{author}{S.D. Leigh}, \bibinfo{author}{M.~Levenson},
  \bibinfo{author}{M.~Vangel}, \bibinfo{author}{D.L. Banks},
  \bibinfo{author}{N.A. Heckert}, \bibinfo{author}{J.F. Dray},
  \bibinfo{author}{S.~Vo}, \bibinfo{title}{{SP 800-22 Rev. 1a. A Statistical
  Test Suite for Random and Pseudorandom Number Generators for Cryptographic
  Applications}}, \bibinfo{type}{Technical Report}, National Institute of
  Standards \& Technology, \bibinfo{address}{Gaithersburg, MD, United States},
  \bibinfo{year}{2010}.
  \bibinfo{note}{\url{https://csrc.nist.gov/projects/random-bit-generation/documentation-and-software}}.
\bibitem[{Coron(1999)}]{Coron1999}
\bibinfo{author}{J.S. Coron}, \bibinfo{title}{{On the Security of Random
  Sources}}, in: \bibinfo{booktitle}{Public Key Cryptography: Second
  International Workshop on Practice and Theory in Public Key Cryptography,
  PKC'99 Kamakura, Japan, March 1--3, 1999 Proceedings},
  \bibinfo{publisher}{Springer Berlin Heidelberg}, \bibinfo{address}{Berlin,
  Heidelberg}, \bibinfo{year}{1999}, pp. \bibinfo{pages}{29--42}.
\bibitem[{Fishman(1978)}]{Fishman:1978:PDE:539984}
\bibinfo{author}{G.S. Fishman}, \bibinfo{title}{{Principles of Discrete Event
  Simulation}}, \bibinfo{publisher}{John Wiley \& Sons, Inc.},
  \bibinfo{address}{New York, NY, USA}, \bibinfo{year}{1978}.
\bibitem[{Hamano and Kaneko(2007)}]{Hamano:2007:COT:1521215.1521232}
\bibinfo{author}{K.~Hamano}, \bibinfo{author}{T.~Kaneko},
  \bibinfo{title}{{Correction of Overlapping Template Matching Test Included in
  {NIST} Randomness Test Suite}}, \bibinfo{journal}{IEICE Trans. Fundam.
  Electron. Commun. Comput. Sci.} \bibinfo{volume}{E90-A}
  (\bibinfo{year}{2007}) \bibinfo{pages}{1788--1792}.
\bibitem[{Kim et~al.(2004)Kim, Umeno and Hasegawa}]{journals/iacr/KimUH04}
\bibinfo{author}{S.J. Kim}, \bibinfo{author}{K.~Umeno},
  \bibinfo{author}{A.~Hasegawa}, \bibinfo{title}{{Corrections of the NIST
  Statistical Test Suite for Randomness}}, \bibinfo{journal}{IACR Cryptology
  ePrint Archive} \bibinfo{volume}{2004} (\bibinfo{year}{2004})
  \bibinfo{pages}{18}.
\bibitem[{Knuth(1997)}]{Knuth:1997:ACP:270146}
\bibinfo{author}{D.E. Knuth}, \bibinfo{title}{{The art of computer programming,
  volume 2 (3rd ed.): seminumerical algorithms}},
  \bibinfo{publisher}{Addison-Wesley}, \bibinfo{year}{1997}.
\bibitem[{L'Ecuyer(1992)}]{rLEC92a}
\bibinfo{author}{P.~L'Ecuyer}, \bibinfo{title}{{Testing Random Number
  Generators}}, in: \bibinfo{booktitle}{Proceedings of the 1992 Winter
  Simulation Conference}, \bibinfo{publisher}{{IEEE} Press},
  \bibinfo{year}{1992}, pp. \bibinfo{pages}{305--313}.
\bibitem[{L'Ecuyer(1994)}]{MR1310607}
\bibinfo{author}{P.~L'Ecuyer}, \bibinfo{title}{{Uniform random number
  generation}}, \bibinfo{journal}{{Annals of Operations Research}}
  \bibinfo{volume}{53} (\bibinfo{year}{1994}) \bibinfo{pages}{77--120}.
\bibitem[{L'Ecuyer(1997)}]{doi:10.1080/00949659708811859}
\bibinfo{author}{P.~L'Ecuyer}, \bibinfo{title}{{Tests Based on Sum-Functions of
  Spacings for Uniform Random Numbers}}, \bibinfo{journal}{Journal of
  Statistical Computation and Simulation} \bibinfo{volume}{59}
  (\bibinfo{year}{1997}) \bibinfo{pages}{251--269}.
\bibitem[{L'Ecuyer et~al.(2000)L'Ecuyer, Cordeau and
  Simard}]{doi:10.1287/opre.48.2.308.12385}
\bibinfo{author}{P.~L'Ecuyer}, \bibinfo{author}{J.F. Cordeau},
  \bibinfo{author}{R.~Simard}, \bibinfo{title}{{Close-Point Spatial Tests and
  Their Application to Random Number Generators}}, \bibinfo{journal}{Operations
  Research} \bibinfo{volume}{48} (\bibinfo{year}{2000})
  \bibinfo{pages}{308--317}.
\bibitem[{L'Ecuyer and Simard(2007)}]{L'Ecuyer:2007:TCL:1268776.1268777}
\bibinfo{author}{P.~L'Ecuyer}, \bibinfo{author}{R.~Simard},
  \bibinfo{title}{{TestU01: A C Library for Empirical Testing of Random Number
  Generators}}, \bibinfo{journal}{ACM Trans. Math. Softw.} \bibinfo{volume}{33}
  (\bibinfo{year}{2007}) \bibinfo{pages}{22:1--22:40}.
\bibitem[{L'Ecuyer and Simard(2013)}]{TestU01Manual}
\bibinfo{author}{P.~L'Ecuyer}, \bibinfo{author}{R.~Simard},
  \bibinfo{title}{{T}est{U}01: {A} {S}oftware {L}ibrary in {ANSI} {C} for
  {E}mpirical {T}esting of {R}andom {N}umber {G}enerators, {S}oftware user's
  guide, version of {M}ay 16},
  \bibinfo{howpublished}{\url{http://simul.iro.umontreal.ca/testu01/tu01.html}},
  \bibinfo{year}{2013}.
\bibitem[{L'Ecuyer et~al.(2002)L'Ecuyer, Simard and Wegenkittl}]{rLEC02c}
\bibinfo{author}{P.~L'Ecuyer}, \bibinfo{author}{R.~Simard},
  \bibinfo{author}{S.~Wegenkittl}, \bibinfo{title}{{Sparse Serial Tests of
  Uniformity for Random Number Generators}}, \bibinfo{journal}{{SIAM} Journal
  on Scientific Computing} \bibinfo{volume}{24} (\bibinfo{year}{2002})
  \bibinfo{pages}{652--668}.
\bibitem[{Marsaglia(1985)}]{1676623w}
\bibinfo{author}{G.~Marsaglia}, \bibinfo{title}{{Note on a Proposed Test for
  Random Number Generators}}, \bibinfo{journal}{IEEE Transactions on Computers}
  \bibinfo{volume}{C-34} (\bibinfo{year}{1985}) \bibinfo{pages}{756--758}.
\bibitem[{Matsumoto and Nishimura(1998)}]{DBLP:journals/tomacs/MatsumotoN98}
\bibinfo{author}{M.~Matsumoto}, \bibinfo{author}{T.~Nishimura},
  \bibinfo{title}{{Mersenne Twister: A 623-Dimensionally Equidistributed
  Uniform Pseudo-Random Number Generator}}, \bibinfo{journal}{ACM Trans. Model.
  Comput. Simul.} \bibinfo{volume}{8} (\bibinfo{year}{1998})
  \bibinfo{pages}{3--30}.
\bibitem[{Okutomi and Nakamura(2010)}]{110007504717}
\bibinfo{author}{H.~Okutomi}, \bibinfo{author}{K.~Nakamura}, \bibinfo{title}{{A
  Study on Rational Judgement Method of Randomness Property Using {NIST}
  Randomnes Test (in {J}apanese)}}, \bibinfo{journal}{The transactions of the
  Institute of Electronics, Information and Communication Engineers. A}
  \bibinfo{volume}{93} (\bibinfo{year}{2010}) \bibinfo{pages}{11--22}.
\bibitem[{Pareschi et~al.(2012)Pareschi, Rovatti and Setti}]{6135498}
\bibinfo{author}{F.~Pareschi}, \bibinfo{author}{R.~Rovatti},
  \bibinfo{author}{G.~Setti}, \bibinfo{title}{{On Statistical Tests for
  Randomness Included in the NIST SP800-22 Test Suite and Based on the Binomial
  Distribution}}, \bibinfo{journal}{IEEE Transactions on Information Forensics
  and Security} \bibinfo{volume}{7} (\bibinfo{year}{2012})
  \bibinfo{pages}{491--505}.
\bibitem[{Simard and L'Ecuyer(2011)}]{JSSv039i11}
\bibinfo{author}{R.~Simard}, \bibinfo{author}{P.~L'Ecuyer},
  \bibinfo{title}{{Computing the Two-Sided Kolmogorov-Smirnov Distribution}},
  \bibinfo{journal}{Journal of Statistical Software, Articles}
  \bibinfo{volume}{39} (\bibinfo{year}{2011}) \bibinfo{pages}{1--18}.

\end{thebibliography}





\end{document}